\documentclass[amssymb, nobibnotes, aps,pre,twocolumn]{revtex4-1}
\usepackage{graphicx}
\usepackage{amsmath}
\usepackage{amssymb}
\usepackage{epsfig,epstopdf}
\usepackage{bm,mathrsfs}
\usepackage{threeparttable}
\usepackage{comment}
\usepackage{ulem}
\usepackage[large]{subfigure}
\usepackage{color}
\usepackage[top=30truemm,bottom=30truemm,left=25truemm,right=25truemm]{geometry}
\begin{document}
%

\title{Efficiency at Maximum Power of Laser Quantum Heat Engine Enhanced by Noise-Induced Coherence}

\author{Konstantin E. Dorfman}
\email{Email: dorfmank@lps.ecnu.edu.cn}
\affiliation{State Key Laboratory of Precision Spectroscopy, East China Normal University, Shanghai 200062, China}

\author{Dazhi Xu}
\affiliation{Department of Physics and Center for Quantum Technology Research, Beijing Institute of Technology,5 South Zhongguancun Street, Beijing 100081, China}

\author{Jianshu Cao}
\affiliation{Department of Chemistry, Massachusetts Institute of Technology, Cambridge, MA02139, USA}
\affiliation{
Beijing Computational Science Research Center, Beijing 100084, China}

\date{\today}


\affiliation{}

\begin{abstract}
	Quantum coherence has been demonstrated in various systems including organic solar cells and solid state devices. In this letter, we report the lower and upper bounds for the performance of quantum heat engines determined by the efficiency at maximum power. Our prediction based on the canonical 3-level Scovil and Schulz-Dubois maser model strongly depends on the ratio of system-bath couplings for the hot and cold baths and recovers the theoretical bounds established previously for the Carnot engine.  Further,  introducing a 4-th level to the maser model can enhance the maximal power and its efficiency, thus demonstrating the importance of quantum coherence in the thermodynamics and operation of the heat engines beyond the classical limit.
\end{abstract}

\maketitle

\section{Introduction}

Systematic formulation of the laws of thermodynamics has been entangled with the notion of heat engines. The key elements of a heat engine include \cite{van94} the working fluid (e.g. steam), two-temperature environment (thermal bath and the entropy sink), and net work along with relevant measurement in the form of the output power. A quantum heat engine (QHE) can be viewed as a miniature version of the classical heat engine on the scale where quantum effects cannot be neglected \cite{ben17}. Examples include lasers, solar cells, photosynthetic organisms, etc. The basic ``quantum'' feature of these QHEs is rooted in the fact that the working fluid is a few-level quantum system \cite{scu03,kos14,jar16}. Quantum coherence has been identified as an important performance enhancement feature of continuous devices \cite{che16,whi16,aga17}. The sources of ``quantumness'' may also include nonclassical baths \cite{ros14,ali15}, quantum feedback \cite{um15}, quantum measurements\cite{hay17,elo17}, or quantum effects in the interaction with environment \cite{Scully:2003cd,li14,kil15,thi16}. The latter effects have been explored in the context of quantum coherence in system-bath interactions. In particular, it has been shown that noise-induced coherence may enhance power of the laser and solar cell \cite{scu11,yao15} and is responsible for highly efficient energy transfer in photosynthetic systems \cite{Dorfman:2013PNAS}, which has been confirmed in the experimental studies of polymer solar cells \cite{bit14} and recently demonstrated in nitrogen vacancy-based microscopic QHE \cite{kla17} in diamonds and generalized to any kind of quantum effects \cite{obi14,kla17}. However, the demonstrated power enhancement occurs in the limit of low efficiency and, therefore, the practical implementation of the effect might be difficult. Furthermore, unlike the power the efficiency itself usually shows no sign of quantum coherence\cite{}. Therefore, although it is expected that efficiency at maximum power will be affected by coherence primarily due to the power dependence, the exact analytical form and the range for efficiency at maximum power analogous to that of a traditional QHE \cite{2010PhRvL.105o0603E} has not been achieved so far.

Here we calculate the fundamental bound for the efficiency at maximum power (EMP) \cite{wan12} with and without coherence. We demonstrate that the Chambadal-Novikov-Curzon-Ahlborn (CNCA) limit \cite{nov57,cham57,1975AmJPh} is not a fundamental bound but rather the result of a particular parameter optimization, which has been shown in stochastic engines \cite{sch07,ben17}. We further derive analytical expressions for both the high and low temperature regimes and demonstrate that coherence may indeed increase both the EMP and power. Finally, we obtain a strong indication that a broken symmetry can affect the EMP depending on the relation between the couplings of the nondegenerate states responsible for coherence. These results suggest a robust mechanism of improving and controlling the operation of the QHE by manipulation of quantum coherence in the system-environment interaction.

\section{3-level QHE model}
\begin{figure*}[t]
\begin{center}
\includegraphics[trim=0cm 0cm 0cm 0cm, angle=0, width=0.9\textwidth]{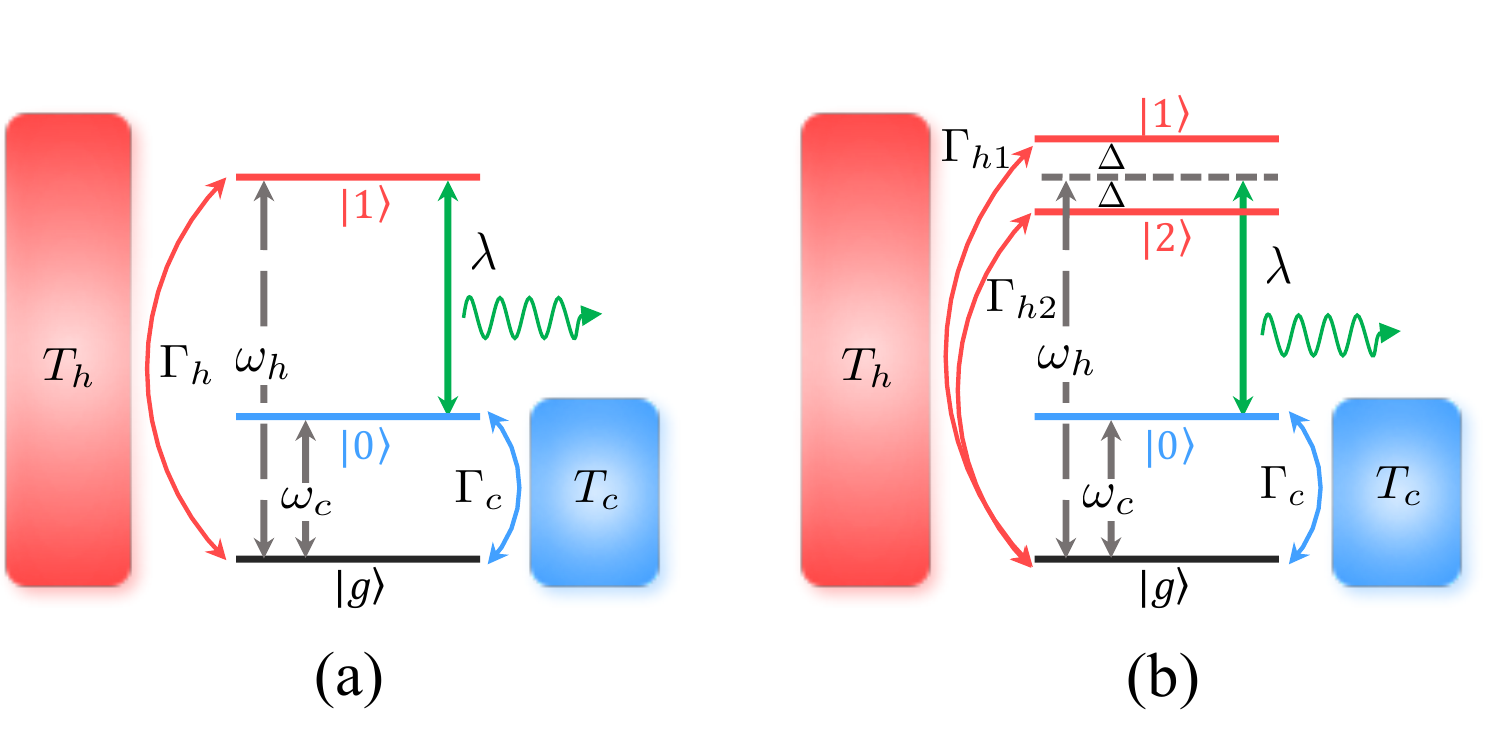}
\end{center}
\caption{(Color online)(a) Three-level QHE model. The hot (cold) reservoir is coupled with the $\vert1\rangle-\vert g\rangle$ ($\vert0\rangle-\vert g\rangle$) transition with dissipative rate $\Gamma_{h}$ ($\Gamma_{c}$). The single mode laser field is driving between $\vert1\rangle$ and $\vert 0\rangle$ with coupling strength $\lambda$. (b) Four-level QHE model. The difference from three-level model is that there are two excited states $\vert 1\rangle$ and $\vert 2\rangle$ coupled with $\vert 0\rangle$ via the hot reservoir, with dissipative rate $\Gamma_{h1}$ and $\Gamma_{h2}$ respectively. The energy gap between $\vert 1\rangle$ and $\vert 2\rangle$ is $2\Delta$.}
\label{fig:scheme}
\end{figure*}

 The model is represented by a three-level system - ground state $\vert g\rangle$, and two excited states $\vert1\rangle$ and $\vert0\rangle$ that correspond to the lasing transition in a Scovil Schulz-DuBois (SSD) laser \cite{Scovil:1959gt}. A hot reservoir with temperature $T_{h}$ drives the $\vert1\rangle-\vert g\rangle$ transition, whereas a cold reservoir with temperature $T_{c}$ is coupled to the $\vert0\rangle-\vert g\rangle$ transition (see Fig. \ref{fig:scheme}a). In the rotating frame an arbitrary operator is defined as $A_R=e^{\frac{i}{\hbar}\bar{H}t}A_Se^{-\frac{i}{\hbar}\bar{H}t}$ where $A_S$ is the operator in the Schr\"{o}dinger picture and $\bar{H}=\hbar\omega_g|g\rangle\langle g|+\frac{\hbar\omega}{2}|1\rangle\langle 1|-\frac{\hbar\omega}{2}|0\rangle\langle 0|$ with $\hbar\omega_g$ the ground state energy and $\omega$ the laser frequency. The time evolution of the system density matrix in the rotating frame \cite{bou07} is described by
\begin{align}\label{eq:DM1}
\dot{\rho}_R=-\frac{i}{\hbar}[H_0-\bar{H}+V_{R},\rho_R]+\mathcal{L}_c[\rho_R]+\mathcal{L}_h[\rho_R],
\end{align}
where the Hamiltonian of the system is given by $H_0=\hbar\sum_{i=g,0,1}\omega_i|i\rangle\langle i|$, and $\omega_{i}$ represents the relevant energy. Interaction with the single mode lasing field is described by semiclassical Hamiltonian
\begin{align}\label{eq:VR1}
V_R=\hbar\lambda(|1\rangle\langle 0|+|0\rangle\langle 1|),
\end{align}
where $\lambda$ is the field-matter coupling, which is considered to be strong compared to any other relaxation process. System-bath interaction is described by a Liouvillian $\mathcal{L}_j[\rho]=\Gamma_j(n_j+1)[2|g\rangle\langle g|\rho_{\nu_j\nu_j}-|\nu_j\rangle\langle\nu_ j|\rho-\rho|\nu_j\rangle\langle \nu_j|]+\Gamma_jn_j[2|\nu_j\rangle\langle\nu_j|\rho_{gg}-|g\rangle\langle g|\rho-\rho|g\rangle\langle g|]$, $j=c,h$, $\nu_c=0$, $\nu_h=1$ (see Appendix \ref{sec:SSD}). Here the average occupation numbers are given by 
$n_c=\left(\exp[\hbar\omega_{c}/(k_BT_c)]-1\right)^{-1}$,$n_h=\left(\exp[\hbar\omega_{h}/(k_BT_h)]-1\right)^{-1}$ with $\omega_c=\omega_{0}-\omega_{g}$, $\omega_h=\omega_{1}-\omega_{g}$.

\begin{figure}[t]
\begin{center}
\includegraphics[trim=0cm 0cm 0cm 0cm, angle=0, width=0.45\textwidth]{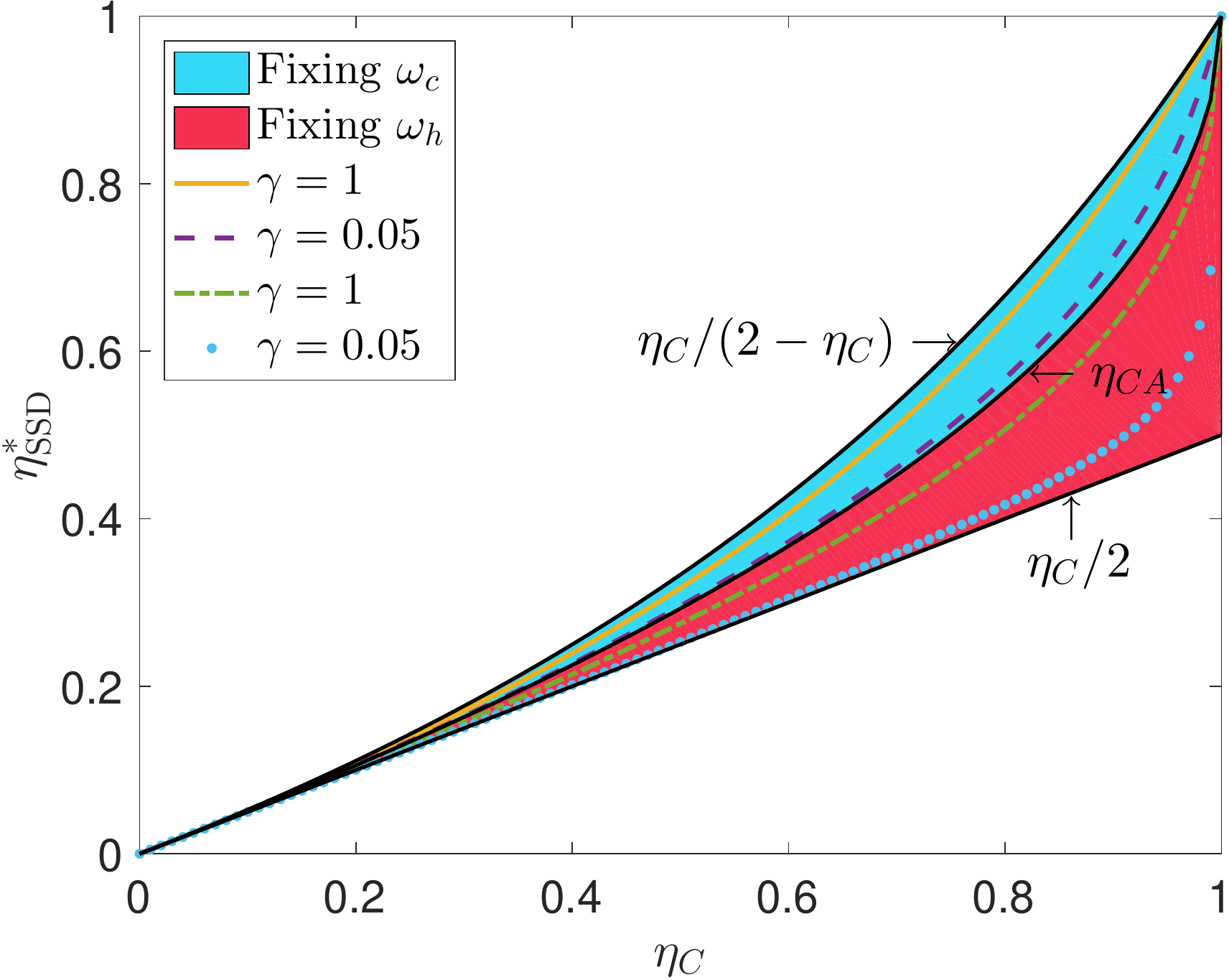}
\end{center}
\caption{(Color online) EMP $\eta_{\mathrm{SSD}}^{\ast}$ vs the Carnot efficiency $\eta_{C}$ for the three-level QHE model. When fixing $\omega_{h}$ and optimizing $\omega_c$, $\eta_{\mathrm{SSD}}^{\ast}$ lies between the bounds $\eta_{C}/2$ and $\eta_{\mathrm{CA}}$ (red area). When fixing $\omega_{c}$ and optimizing $\omega_h$, $\eta_{\mathrm{SSD}}^{\ast}$ lies between the bounds $\eta_{\mathrm{CA}}$ and $\eta_{C}/(2-\eta_{C})$ (blue area). $\eta_{\mathrm{SSD}}^{\ast}$ approaches the lower (upper) bound when $\gamma\rightarrow 0(\infty)$. We present the $\eta_{\mathrm{SSD}}^{\ast}$ with $\gamma=1$ (green dash-dotted line) and $\gamma=0.05$ (blue dotted line) fixing $\omega_h$; $\gamma=1$ (yellow solid line) and $\gamma=0.05$ (purple dash line) fixing $\omega_c$. Here we set $k_{\mathrm{B}}T_{h}=100\Gamma_c$ and $\lambda=1000\Gamma_{c}$.}
\label{fig:etssg}
\end{figure}

Following the general approach outlined in Ref. \cite{bou06,bou07} that is applicable for weak system-bath coupling, the power, heat flux, and efficiency of the QHE are defined, respectively, as
\begin{align}\label{eq:Pdef}
P=-\frac{i}{\hbar}\text{Tr}\left([H_0,V_R]\rho_R\right),
\end{align}
\begin{align}\label{eq:Qhdef}
\dot{Q}_h=\text{Tr}\left(\mathcal{L}_h[\rho_R]H_0\right),
\end{align}
\begin{align}\label{eq:etdef}
\eta=-\frac{P}{\dot{Q}_h}.
\end{align}
In the steady state one can calculate Eqs. (\ref{eq:Pdef}) - (\ref{eq:etdef}) and obtain the efficiency $\eta_{\mathrm{SSD}}=1-\omega_{c}/\omega_{h}$. The maximum possible efficiency is given by the Carnot efficiency, which can be shown by equating lasing level populations using  equilibrium distributions:  $\rho_{11}/\rho_{gg}=\exp[{-\hbar\omega_{h}/(k_{\mathrm{B}}T_h)}]$ and $\rho_{00}/\rho_{gg}=\exp[{-\hbar\omega_{c}/(k_{\mathrm{B}}T_c)}]$. This yields maximum efficiency to be Carnot $\eta_{\mathrm{SSD}}<\eta_C\equiv 1-T_c/T_h$. However, the QHE with the Carnot efficiency has zero power as the Carnot limit corresponds to the threshold condition for the lasing, rather than the optimum operation with the highest power. Here we first maximize the power and then calculate the corresponding EMP $\eta^{*}\equiv \eta(P_{\text{max}})$.  The power (\ref{eq:Pdef}) and efficiency (\ref{eq:etdef}) in the high temperature limit can be calculated by setting $n_h\simeq k_{\mathrm{B}}T_h/(\hbar\omega_{h})$ and $n_c\simeq k_{\mathrm{B}}T_c/(\hbar\omega_{c})$ and assuming the strong coupling limit $\lambda\gg \Gamma_{h,c}$. The maximization of the power is typically performed  for the parameter $c=\omega_h/\omega_c$ with respect to the temperature ratio $\tau=T_c/T_h$ \cite{kos14}. One way to perform this optimization is to fix $\omega_h$ while varying $\omega_c=\omega_h/c$ to obtain the maximum power. More details regarding the optimization of the output power are presented in Appendix \ref{sec:CA}, giving the corresponding efficiency
\begin{align}\label{eq:etssdwh}
&\eta_{\mathrm{SSD}}^{(\omega_h)*}=\gamma^{-1}[\tau+\gamma-\sqrt{\tau(1+\gamma)(\tau+\gamma)}],
\end{align}
where $\gamma=\Gamma_h/\Gamma_c$. Alternatively one can fix $\omega_c$ and vary $\omega_h=c\omega_c$ which yields
\begin{align}\label{eq:etssdwc}
\eta_{\mathrm{SSD}}^{(\omega_c)*}=1-\frac{\tau}{\sqrt{(1+\gamma)(\tau+\gamma)}-\gamma}.
\end{align}
Note that the difference in the two optimization schemes have an underlying physical reason. 
Since evolution due to the coupling to cold and hot baths do not commute,
the absence of the time translational invariance results in different operations under two limiting conditions.
In Eqs. (\ref{eq:etssdwh})-(\ref{eq:etssdwc}) the superscript $\omega_j$ $(j=h,c)$ denotes the fixed parameter. Further optimization involves the parameter $\gamma$. We obtain the lower bound for the efficiency at $\gamma\to 0$  and the upper bound at $\gamma\to\infty$:
\begin{align}\label{eq:etwhboundary}
(1-\tau)/2\leq \eta_{\mathrm{SSD}}^{(\omega_h)*}\leq 1-\sqrt{\tau},
\end{align}
\begin{align}\label{eq:etwcboundary}
1-\sqrt{\tau}\leq \eta_{\mathrm{SSD}}^{(\omega_c)*}\leq \frac{1-\tau}{1+\tau}.
\end{align}
Note, that $\gamma\to\infty$ limit should be achieved while keeping $\Gamma_h\ll\omega_h$ and $\Gamma_c\ll\omega_c$ ensuring the weak dissipation regime. Both the lower bound for $\eta_{\mathrm{SSD}}^{(\omega_c)*}$ and the upper bound for $\eta_{\mathrm{SSD}}^{(\omega_h)*}$ are given by $\eta_{\mathrm{CA}}=1-\sqrt{T_c/T_h}$, which is known as CNCA efficiency \cite{nov57,cham57,1975AmJPh}. As show in Fig. \ref{fig:etssg},  $\eta_{\mathrm{CA}}$ separates the entire parameter regime of $\eta_{\mathrm{SSD}}^{\ast}$ into two parts; one corresponds to fixing $\omega_{h}$ (red area) with upper bound $\eta_{\mathrm{CA}}$ and lower bound $\eta_{\mathrm{C}}/2$, and the other one corresponds to fixing $\omega_{c}$ (blue area) with upper bound $\eta_{\mathrm{C}}/(2-\eta_{\mathrm{C}})$ and lower bound $\eta_{\mathrm{CA}}$. We also present the numerical simulation for $\gamma=1$ and $\gamma=0.05$ for the two different optimization approaches, which shows that the larger (smaller) $\gamma$ is, the closer $\eta_{\mathrm{SSD}}^{\ast}$ approaches to the upper (lower) bound. Our result is therefore agrees with the work of Esposito et al \cite{2010PhRvL.105o0603E}, which shows that the CNCA is not a fundamental limit for the QHE performance but rather a consequence of the optimization procedure. However, despite of the mathematical equivalence between the two results,  the physics explored here is rather different. The previous analysis of EMP bounds is based on the 4-stroke engine and therefore employs finite time expansion.  In contrast, this paper analyzes continuous QHE and therefore uses the steady-state solution of quantum master equation, which is a non-perturbative treatment of thermodynamics. 	Note also that the CNCA limit in Ref. \cite{2010PhRvL.105o0603E} is achieved when $\gamma=1$, which corresponds to the symmetric case when the system dynamics during the interaction with cold and hot bath commute. This holds only for the linear dependence of heat flux with respect to $\gamma_h$ and $\gamma_c$, which applies in the weak dissipation limit. Our calculation shows that the CNCA is achieved in the $\gamma\to 0$ or $\gamma\to \infty$ regime, depending on the optimization scheme. While our method is based on master equation approach which is second-order to system-bath coupling, the solution for the density operator as well as all consequent calculations of the heat flux, entropy and power are performed exactly to all orders in the system-bath coupling. In that way this is applicable for a wide range of parameters and does not assume a particular form of the dependence of the thermodynamical quantities with respect to system-bath coupling. 

We also note, that $\eta_{\mathrm{SSD}}^{(\omega_h)*}$ and $\eta_{\mathrm{SSD}}^{(\omega_c)*}$ are analogous to classical quantities $\eta_{-}$ and $\eta_{+}$ discussed in Ref. \cite{joh16}, which are obtained for the systems with power law dependence of the internal energy with respect to entropy $U=S^\omega$. This parameter $\omega$ is analogous to $c=\omega_h/\omega_c$ in the present case. Eqs. (\ref{eq:Ph}) and (\ref{eq:Pc}) predict similar result for the QHE power $P\sim c^\omega$ with the dominant contributions in the range $1<\omega<2$ which includes limits such as black-body radiation (phonon bath) regime \cite{joh16}. Indeed, the phonon bath in the high temperature limit yields classical result.

\section{4-level QHE model}
\begin{figure*}[t]
\begin{center}
\includegraphics[trim=0cm 0cm 0cm 0cm, angle=0, width=0.78\textwidth]{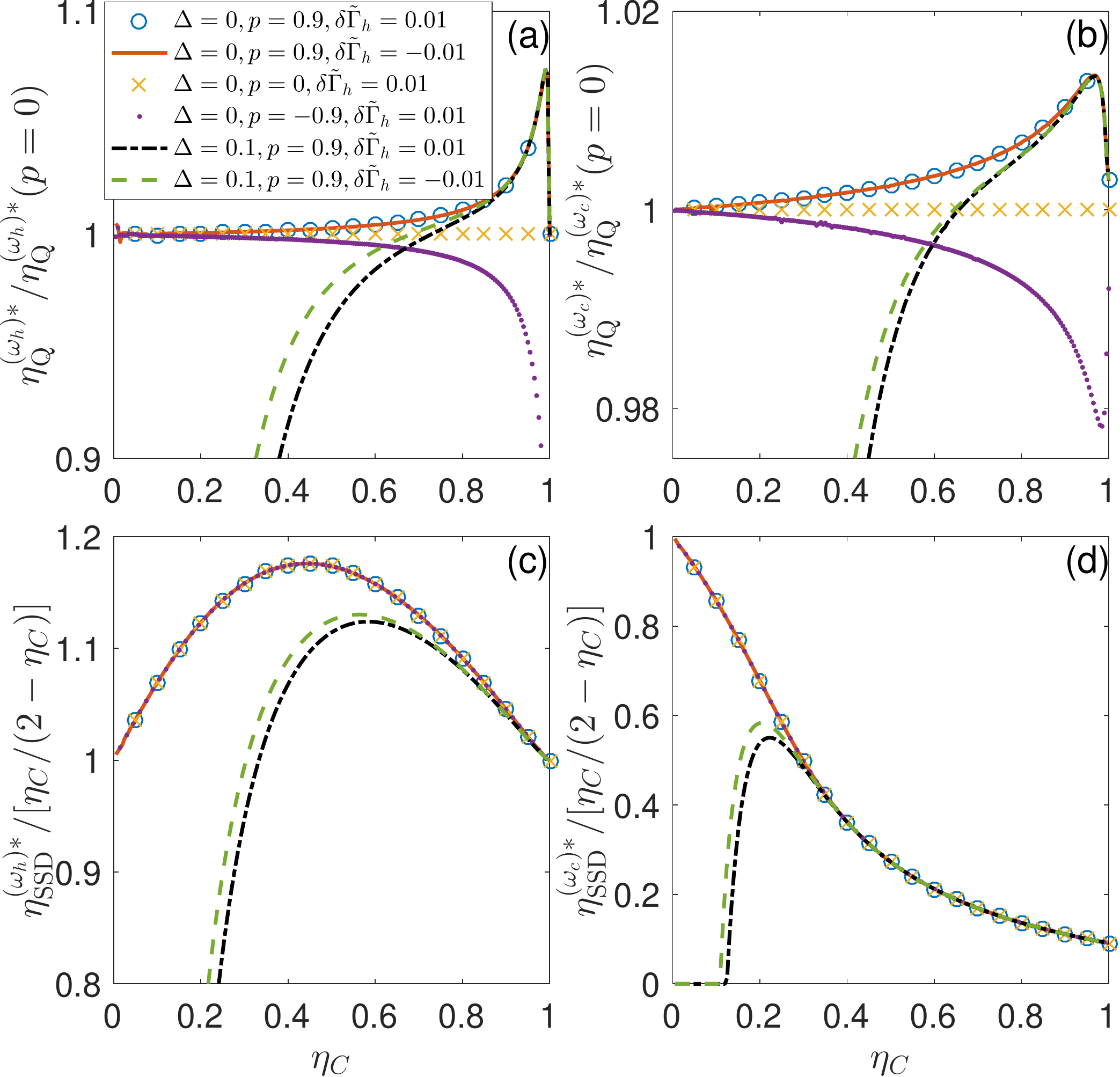}
\end{center}
\caption{(Color online) EMP vs the Carnot efficiency of the four-level QHE model in the high temperature for (a) fixing $\omega_{h}$, (b) fixing $\omega_{c}$; and in the low temperature for (c) fixing $\omega_h$ and (d) fixing $\omega_c$. In the high temperature we normalize the plots with respect to EMP at $p=0$, and set $k_{\mathrm{B}}T_h=100\Gamma_c$, $\lambda=1000\Gamma_c$; in the low temperature we normalize the plots with respect to  $\eta_{C}/(2-\eta_{C})$, and set $k_{\mathrm{B}}T_h=0.1\Gamma_c$, $\lambda=10\Gamma_c$. In the degenerate case $\Delta=0$, the EMP is enhanced by increasing the coherence $p$ when $T_h$ is high, while the coherence does not affect the EMP when $T_h$ is low. 
Interchanging $\Gamma_{h1}$ and $\Gamma_{h2}$ ($\delta\tilde{\Gamma}_h\equiv(\Gamma_{h1}-\Gamma_{h2})/\Gamma_c=\pm0.01$) gives the same result (blue circle line and red solid line) for $\Delta=0$. While for $\Delta=0.1$, the EMP of $\delta\tilde{\Gamma}_h=\pm0.01$ (black dash-dotted line and green dash line) show obvious differences and are lower than their degenerate counterparts especially for small $\eta_C$.}
\label{fig:4level}
\end{figure*}

We now consider a coherence-enhanced QHE model where we replace a single level $\vert1\rangle$ in the SSD model by a pair of closely spaced states $\vert1\rangle$ and $\vert2\rangle$ \cite{tsc14} separated by $2\Delta$ (see Fig. \ref{fig:scheme}b). The interaction Hamiltonian with lasing radiation, in the rotating frame  Eq. (\ref{eq:VR1}) can be then recast as
\begin{align}\label{eq:VR2}
V_R=\hbar\lambda(|1\rangle\langle 0|+|2\rangle\langle 0|+|0\rangle\langle 1|+|0\rangle\langle 2|).
\end{align}
Density matrix evolution can be described by Eq. (\ref{eq:DM1}) by replacing $\bar{H}\to\bar{\bar{H}}$ with $\bar{\bar{H}}=\hbar\omega_g|g\rangle\langle g|+\frac{\hbar\omega}{2}(|1\rangle\langle 1|+|2\rangle\langle 2|-|0\rangle\langle 0|)$, and the new bath Liouvillian is given by Eq. (\ref{eq:Lh1}). The coupling of states $\vert1\rangle$ and $\vert2\rangle$ to the ground state is governed by spontaneous emission rates $\Gamma_{h1}$ and $\Gamma_{h2}$, respectively, and thermal radiation occpuation numbers $n_{h1,2}=\left(\exp\left[\hbar(\omega_{h}\pm \Delta)/(k_BT_h)\right]-1\right)^{-1}$. The details of the time evolution of the density matrix equations are given in Appendix \ref{sec:DMcoh}. The key parameter that characterizes the strength of the noise-induced coherence between states $\vert1\rangle$ and $\vert2\rangle$ is  the dipole alignment factor $p=\frac{\mathbf{\mu_{1g}}\cdot\mathbf{\mu_{2g}}}{|\mathbf{\mu_{1g}}||\mathbf{\mu_{2g}}|}$: $p=0$ yields no hot bath coherence, whereas $p=1$ ($p=-1$) corresponds to maximum coherence due to constructive (destructive) interference. Assuming that the laser frequency is tuned midway between states $\vert1\rangle$ and $\vert2\rangle$: $\omega=(\omega_{10}+\omega_{20})/2=\omega_{10}-\Delta=\omega_{20}+\Delta$, the efficiency of the quantum-enhanced QHE is given by
\begin{align}\label{eq:etQ}
\eta_Q=1-\frac{\omega_{c}}{\omega_h+\Delta\xi},
\end{align}
where subscript $Q$ signifies quantum enhancement in the four-level system, $\omega_h=\omega+\omega_c$ and $\xi=(\rho_{01}-\rho_{10}-\rho_{02}+\rho_{20})/(\rho_{01}-\rho_{10}+\rho_{02}-\rho_{20})$. Depending on the sign of $\xi$ one can get either enhancement ($\xi>0$) or suppression ($\xi<0$) of the efficiency. There are two sources of the quantum coherence in the system. First is the laser field, which gives a rise to the coupling between $\rho_{10}$ and $\rho_{12}$ in Eq. (\ref{eq:r10}). Due to strong laser-system coupling, this contribution can be quite substantial. The second one is due to quantum interference in the hot bath and depends on the magnitude of $p$. In the high temperature limit for degenerate states $\Delta=0$ and $n_{h1}=n_{h2}=n_h$ we calculate the EMP, as shown in Figs. \ref{fig:4level}a-\ref{fig:4level}b, and obtain the same form of Eqs. (\ref{eq:etssdwh}) and  (\ref{eq:etssdwc}) by fixing $\omega_h$ and  $\omega_c$, respectively. The parameter $\gamma$ is now replaced by $\gamma_p=(\Gamma_{h1}+\Gamma_{h2}+2p\sqrt{\Gamma_{h1}\Gamma_{h2}})/(2\Gamma_c)$, which depends on the $p$, such that $\eta_{\mathrm{SSD}}(\gamma)\to\eta_Q(\gamma_p)$. Note that for $\Gamma_{h1}=\Gamma_{h2}$, $\gamma_{p=0}=\gamma$ gives exactly the same efficiency as  for the three-level system. However, for $\gamma_{p=-1}=0$ and $\gamma_{p=1}=2\gamma$, Eqs. (\ref{eq:etssdwh}) - (\ref{eq:etssdwc}) yield a very different result. In addition for $\gamma_p\to 0$, we obtain
\begin{align}\label{eq:efwhr0}
\eta_{Q}^{(\omega_h)*}\simeq\frac{1-\tau}{2}+\frac{(1-\tau)^2\gamma_p^2}{16t},
\end{align}
and
\begin{align}\label{eq:efwcr0}
\eta_{Q}^{(\omega_c)*}\simeq1-\sqrt{\tau}+\frac{(1-\sqrt{\tau})^2}{4\sqrt{\tau}}\gamma_p^2.
\end{align}
Similarly for $\gamma_p\to\infty$ we obtain
\begin{align}\label{eq:efwhrinf}
\eta_{Q}^{(\omega_h)*}\simeq 1-\sqrt{\tau}-\frac{\sqrt{\tau}}{\gamma_p^2}(1-\sqrt{\tau})^2,
\end{align}
and
\begin{align}\label{eq:efwcrinf}
\eta_{Q}^{(\omega_c)*}(\gamma_p\to\infty)\simeq \frac{1-\tau}{1+\tau}-\frac{\tau(1-\tau)^2}{\gamma_p^2(1+\tau)^2}.
\end{align}
One can see that in both limits of $\eta_{Q}^{*}(\gamma_{-1})<\eta_{Q}^{*}(\gamma_{0})<\eta_{Q}^{*}(\gamma_{1})$ coherence can increase (reduce) the EMP for constructive (destructive) interference, respectively. Constructive interference with zero phase delay between the emission and absorption pathways enhances the overall absorption efficiency, whereas the destructive interference accompanied with the $\pi$ phase-shift reduces the overall absorption efficiency. This point is verified by the numerical simulation, in the case of fixing $\omega_h$ (Fig. \ref{fig:4level}a) and fixing $\omega_c$ (Fig. \ref{fig:4level}b), we have $\eta_{Q}^{*}(\gamma_{-0.9})<\eta_{Q}^{*}(\gamma_{0})<\eta_{Q}^{*}(\gamma_{0.9})$. The effect of coherence is most significant when $\eta_{C}$ is large, while in the limits $\eta_{C}\rightarrow0$ and $\eta_{C}\rightarrow1$, the differences induced by the coherence disappear. The nondegenerate case  of $\Delta\neq 0$ and $\Gamma_{h1}\neq\Gamma_{h2}$ deserves  special consideration.  Eq. (\ref{eq:etQ}) may violate the time translational symmetry and thus increase the efficiency. It is indeed true that the quantum efficiency can increase for nondegenerate systems.\cite{Scully:2003cd} However, this is accompanied by the reduced power, which results in a smaller EMP than the degenerate case (see Fig. \ref{fig:4level}), highlighting the importance of the time translational symmetry in thermodynamics \cite{los15}.

Typically the EMP is calculated in the high temperature limit. The maximum effect of doubling of the power has been achieved at low temperature \cite{scu11,har12}. The low temperature regime has to be carefully analyzed due to quantum fluctuations and other related effects. We first keep $\omega_h$ fixed. Introducing $\alpha_h=\hbar\omega_h/(k_\mathrm{B}T_h)\gg 1$ such that $n_h=\exp(-\alpha_h)$ and $n_c=\exp[-\alpha_h/(c\tau)]$, after optimizing the power with respect to $c$, we obtain for EMP
\begin{align}\label{eq:efwh3lev}
\eta_{\mathrm{SSD}}^{(\omega_h)*}=\frac{\tau}{\alpha_h}\left(\mathcal{P}[1+\alpha_h(1-\tau)/\tau]-1\right),
\end{align}
where $\mathcal{P}(e^x)$ is a product log function which is a principal solution for $x$ in $z=xe^{x}$. Note that efficiency defined in Eq. (\ref{eq:efwh3lev}) does not depend on the $\gamma$, which is a consequence of excited states degeneracy. For $\alpha_h\gg1$, $\mathcal{P}[1+\alpha_h(1-\tau)/\tau]\simeq 1+\alpha_h(1-\tau)/\tau$. We thus obtain $\eta_{\mathrm{SSD}}^{(\omega_c)*}(\alpha_h\to\infty)\to\eta_C$ such that the maximum efficiency is governed by the Carnot efficiency. We now turn to the case when  $\omega_c$ is fixed. Introducing  $\alpha_c=\hbar\omega_c/(k_\mathrm{B}T_c)\gg 1$ such that $n_c=\exp(-\alpha_c)$, $n_h=\exp(-c\alpha_c\tau)$ after optimization with respect to $c$ we obtain for EMP
\begin{align}\label{eq:EflowTwc3lev}
\eta_{\mathrm{SSD}}^{(\omega_c)*}=1-\frac{\alpha_c\tau}{1+\alpha_c\tau-\mathcal{P}(1-\alpha_c(1-\tau))}.
\end{align}
For large $\alpha_c$ one can neglect the $\mathcal{P}$ function. Expanding Eq. (\ref{eq:EflowTwc3lev}) for $\alpha_c\gg 1$ we obtain $\eta_{\mathrm{SSD}}^{(\omega_c)*}(\alpha_c\to\infty)\simeq(\tau \alpha_c)^{-1}$ which is much smaller than the lower bound of $\eta_C/2$. Note that introduction of the quantum coherence between degenerate states $\vert1\rangle$ and $\vert2\rangle$ does not make any significant difference in the efficiency. Eq. (\ref{eq:etQ}) reduces to expression for $\eta_{\mathrm{SSD}}^\ast$ and Eqs. (\ref{eq:efwh3lev}) - (\ref{eq:EflowTwc3lev}) hold in this case and does not depend on coherence since there is no functional dependence on $\gamma_p$. The numerical simulation in the low temperature regime shows clearly that in the degenerate case of $\Delta=0$, the $\eta_{\mathrm{SSD}}^{\ast}$ corresponding to different $p$ have the same magnitudes for both cases of fixing $\omega_h$ (Fig. \ref{fig:4level}c) and  fixing $\omega_c$ (Fig. \ref{fig:4level}d). We also notice that, when fixing $\omega_h$, $\eta_{\mathrm{SSD}}^{\ast}$ can surpass the high temperature upper bound $\eta_C/(2-\eta_C)$ for $0<\eta_C<1$. If the two excited states are non-degenerate $\Delta=0.1$, $\eta_{\mathrm{SSD}}^{\ast}$ are lower than the degenerate case, especially when $\eta_C$ is small, when $\eta_C$ approaches $1$, the degenerate and non-degenerate results are coincide with each other. In Fig. \ref{fig:4level}d when $\eta_C \lesssim 0.1$ the $\eta_{\mathrm{SSD}}^{\ast}$ for non-degenerate case are set to zero, because the optimal power is negative, the model no longer represents a heat engine.  


We have analyzed the impact of the coherence on EMP in both the high-temperature limit, when the effect can be significant, and in the low-temperature limit, where the effect is not present. Coming back to the results of Ref. \cite{scu11} on the maximum effect of coherence on the QHE power, we note the following. At high temperature the lower and upper bounds for the maximum power $\text{max}|P|\equiv P^{(*)}$ is achieved at $r\to0$ and $r\to\infty$, respectively, giving 
\begin{align}\label{eq:Pmaxwh}
\frac{1}{6}\gamma\frac{(1-\tau)^2}{\tau}<\frac{P_{\mathrm{SSD}}^{(\omega_h)*}}{\Gamma_c\omega_h}<\frac{2}{3}(1-\sqrt{\tau})^{2}
\end{align}
and 
\begin{align}\label{eq:Pmaxwc}
\frac{2}{3}\gamma\frac{(1-\sqrt{\tau})^2}{\tau}<\frac{P_{\mathrm{SSD}}^{(\omega_c)*}}{\Gamma_c\omega_c}<\frac{1}{6}\frac{(1-\tau)^2}{\tau}.
\end{align}
For the four-level system, $P_{Q}^{*}$ contains an overall factor of $3/4$ compared to $P_{\mathrm{SSD}}^{*}$ and $\gamma$ has to be replaced by $\gamma_p$ in Eqs. (\ref{eq:Pmaxwh}) - (\ref{eq:Pmaxwc}). Then the upper bound, which is independent of $\gamma$ is reduced in the presence of the fourth level. However, for small $\gamma\ll 1$, we obtain $P_{Q}^{*}(\gamma_p\ll 1)/P_{\mathrm{SSD}}^{*}(\gamma\ll 1)=(4\gamma)/(3\gamma_p)$,
which yields for $p=1$ a 50\% enhancement due to coherence, which has been predicted in Ref. \cite{scu11}. 
More generally at low temperature for $\Gamma_{h1}=\Gamma_{h2}=\Gamma_h$ we obtain $P_{Q}^{*}(\gamma_p)/P_{\mathrm{SSD}}^{*}(\gamma)=(\gamma+1)/(\gamma+\frac{1}{1+p})$,
so that the incoherent case ($p=-1$) we get that both powers are equal, whereas for $p=1$ and $\gamma\ll 1$ we get that $P_{Q}^{*}=2P_{\mathrm{SSD}}^{*}$ - 100\% enhancement that has been considered in Ref. \cite{scu11}. One can also see that in this regime the efficiencies for both systems are equal. In contrast, if $\Gamma_{h1}=\Gamma_h\gg \Gamma_{h2}$, then the ratio becomes $P_{Q}^{*}/P_{\mathrm{SSD}}^{*}=\frac{2\gamma}{\gamma+2}/\frac{2\gamma}{\gamma+1}\simeq 1/2$, which gives half of the power (see Appendix \ref{sec:powerlow}). The physical explanation of the effect is relatively simple. Low absorption cross-section at low temperature results in the low power, which can be doubled due to coherence. This regime, however , is most inefficient. In comparison, the maximum possible quantum efficiency corresponds to the Carnot result, which governs so-called open circuit regime in solar cells where the current approaches zero. It is therefore not possible to observe any effect of coherence on the efficiency in this case.

\section{Conclusions} 

In summary, we calculated EMP for a three-level and four-level laser QHE and obtained analytical expressions for the boundaries for EMP obtained via optimization with respect to system-bath coupling strength. As the result, the CNCA efficiency represents a specific parameter regime rather than a fundamental limit for performance of the QHE. Next, using a four-level model, we showed that coherence can enhance both power and EMP beyond the detailed balance value. The modest enhancement of EMP compared to the significant enhancement (up to 100\%) of the power is a result of weak system-bath coupling, which can be further improved in the strong coupling regime \cite{xu16,new17}. Finally, we demonstrated that, for nondegenerate coherent superposition of states, symmetry breaking yields different EMP when couplings to the hot bath is asymmetric. Therefore, coherence is an important quantum feature that strongly affects the operation of the QHE beyond the classical limit.

\section{Acknowledgements}

K.E.D. is supported by the Zijiang Endowed Young
Scholar Fund and Overseas Expertise Introduction Project for Discipline
Innovation (“111 Project,” B12024). D.X. is supported by National Natural Science Foundation of China (Grant No. 11705008). The work of J.C. is supported by NSF (Grant No. CHE-1112825) and CSRC.

\appendix

\section{Three-level Scovil Schulz-DuBois laser QHE}\label{sec:SSD}

The Hamiltonian of the SSD-QHE system is given by
\begin{align}\label{eq:H00}
	H_0=\hbar\sum_{i=g,0,1}\omega_i|i\rangle\langle i|,
\end{align}
where $\hbar\omega_i$ represents relevant energies. Interaction with single mode lasing field of frequency $\omega$ is described by the semiclassical Hamiltonian
\begin{align}
	V(t)=\hbar\lambda(e^{i\omega t}|1\rangle\langle 0|+e^{-i\omega t}|0\rangle\langle 1|),
\end{align}
where $\lambda$ is a field-matter coupling, which is considered to be strong compared to any other relaxation process due to the cold or hot bath. System-bath evolution is described by a Liouvillian for cold
\begin{align}\label{eq:Lc}
	\mathcal{L}_c[\rho]&=\Gamma_c(n_c+1)[2|g\rangle\langle g|\rho_{00}-|0\rangle\langle 0|\rho-\rho|0\rangle\langle 0|]\notag\\
	&+\Gamma_cn_c[2|0\rangle\langle0|\rho_{gg}-|g\rangle\langle g|\rho-\rho|g\rangle\langle g|],
\end{align}
and hot reservoirs
\begin{align}\label{eq:Lh0}
	\mathcal{L}_h[\rho]&=\Gamma_h(n_h+1)[2|g\rangle\langle g|\rho_{11}-|1\rangle\langle 1|\rho-\rho|1\rangle\langle 1|]\notag\\
	&+\Gamma_hn_h[2|1\rangle\langle 1|\rho_{gg}-|g\rangle\langle g|\rho-\rho|g\rangle\langle g|],
\end{align}
where average occupation numbers are given by 
\begin{align}
	n_c=\left(e^{\frac{\hbar\omega_{c}}{k_BT_c}}-1\right)^{-1},\quad n_h=\left(e^{\frac{\hbar\omega_{h}}{k_BT_h}}-1\right)^{-1},
\end{align}
with $\omega_c=\omega_{0g}$, $\omega_h=\omega_{1g}$, and $T_c(T_h)$ being the temperature of the cold (hot) reservoir.

To remove time dependence from the Liouville operators we rotate the eigenbasis by defining operators in the rotational frame: $A_R=e^{\frac{i}{\hbar}\bar{H}t}Ae^{-\frac{i}{\hbar}\bar{H}t}$ where $\bar{H}=\hbar\omega_g|g\rangle\langle g|+\frac{\hbar\omega}{2}|1\rangle\langle 1|-\frac{\hbar\omega}{2}|0\rangle\langle 0|$. Liouville operators for the system-bath interactions remain unchanged $\mathcal{L}_j[\rho_R]=\mathcal{L}_j[\rho]$, $j=c,h$. Density matrix dynamics in the rotating frame is given by Eq. (\ref{eq:DM1}) where $V_R$ is defined by Eq. (\ref{eq:VR1}). The corresponding time evolution of the density matrix elements reads
\begin{align}\label{eq:r11}
	\dot{\rho}_{11}=i\lambda(\rho_{10}-\rho_{01})-2\Gamma_h[(n_h+1)\rho_{11}-n_h\rho_{gg}],
\end{align}
\begin{align}
	\dot{\rho}_{00}=-i\lambda(\rho_{10}-\rho_{01})-2\Gamma_c[(n_c+1)\rho_{00}-n_c\rho_{gg}],
\end{align}
\begin{align}
	\rho_{gg}=1-\rho_{11}-\rho_{00},
\end{align}
\begin{align}\label{eq:r12}
	\dot{\rho}_{10}=&-[i\Delta+\Gamma_h(n_h+1)+\Gamma_c(n_c+1)]\rho_{10}\notag\\
	&+i\lambda(\rho_{11}-\rho_{00}),
\end{align}
where $\Delta=\omega-\omega_1+\omega_0$ is the laser detuning. We then can solve Eqs. (\ref{eq:r11}) - (\ref{eq:r12}) in the steady state by setting $\frac{d}{dt}=0$.

The power, heat flux, and the efficiency of the QHE are defined in Eqs. (\ref{eq:Pdef}) - (\ref{eq:etdef}).
Calculating the traces, we obtain for Eq. (\ref{eq:Pdef})
\begin{align}\label{eq:Pssd}
	P_{SSD}=i\hbar\lambda\omega_{10}(\rho_{01}-\rho_{10}),
\end{align}
where subscript $SSD$ indicates Scovil Schulz-DuBois. Similarly for the heat flux (\ref{eq:Qhdef}) we obtain
\begin{align}\label{eq:Qh1}
	\dot{Q}_{hSSD}=-\hbar\omega_{h}\left(2\Gamma_h[(n_h+1)\rho_{11}-n_h\rho_{gg}]\right).
\end{align}
Using the steady state condition of $\dot{\rho}_{11}=0$, we recast Eq. (\ref{eq:Qh1}) as
\begin{align}
	\dot{Q}_{hSSD}=i\hbar\omega_{h}\lambda(\rho_{01}-\rho_{10}).
\end{align}
The efficiency (\ref{eq:etdef}) reads
\begin{align}\label{eq:eta1}
	\eta_{SSD}=\frac{\omega_{10}}{\omega_{h}}=1-\frac{\omega_{c}}{\omega_{h}},
\end{align}
which is independent of the density matrix elements and is given by the ratio of the lasing frequency.

\section{Efficiency at maximum power for SSD QHE at high temperature}\label{sec:CA}

\subsection{Lower bound. Fixed $\omega_h$}

After solving density matrix equations Eqs. (\ref{eq:r11}) - (\ref{eq:r12}) in the steady state, we obtain for the power (\ref{eq:Pssd}) in the high temperature limit $k_BT_h\gg \hbar\omega_{h}$ and $k_BT_c\gg\hbar\omega_{c}$, corresponding to $n_h\simeq \frac{k_BT_h}{\hbar\omega_{h}}$ and $n_c\simeq \frac{k_BT_c}{\hbar\omega_{c}}$. Then, fixing $\omega_h$ while varying $\omega_c=\omega_h/c$, we obtain
\begin{align}\label{eq:Ph}
	P_{SSD}^{(\omega_h)}=-\frac{2(1-c)(1-c\tau)}{3c(\gamma+c\tau)}\gamma\hbar\Gamma_c\omega_h,
\end{align}
Finally, optimizing the power with respect to the frequency ratio $c$  we obtain
\begin{align}
	\text{max}(P_{SSD}^{(\omega_h)})=&-\frac{2\hbar\Gamma_c\omega_{h}}{3\gamma}[(1-\tau)(\gamma+2)-2(1+\gamma\notag\\
	&-\sqrt{\tau(1+\gamma)(\tau+\gamma)}]
\end{align}
with the efficiency given by Eq. (\ref{eq:etssdwh}).

\subsection{Upper bound. Fixed $\omega_c$}

If we instead keep $\omega_c$ fixed and vary $\omega_h=c\omega_c$, the expression for the power becomes
\begin{align}\label{eq:Pc}
	P_{SSD}^{(\omega_c)}=-\frac{2(1-c)(1-c\tau)}{3(\gamma+c\tau)}\gamma\hbar\Gamma_c\omega_c.
\end{align}
Optimizing the power with respect to the frequency ratio $c$,  we obtain
\begin{align}
	\text{max}(P_{SSD}^{(\omega_c)})=&-\frac{2\gamma\Gamma_c\hbar\omega_c}{3}\frac{(\sqrt{1+\gamma}-\sqrt{\tau+\gamma})^2}{\tau}
\end{align}
with the efficiency given by Eq. (\ref{eq:etssdwc}).

\section{Quantum coherence enhanced laser QHE}\label{sec:DMcoh}

Consider a model when we replace a single level $1$ in the SSD model by a pair of two closely spaced states $1$ and $2$. Eqs. (\ref{eq:H00}) - (\ref{eq:Lh0}) can be then recast as
\begin{align}\label{eq:H01}
	H_0=\hbar\sum_{i=g,0,1,2}\omega_i|i\rangle\langle i|,
\end{align}
\begin{align}
	V(t)=\hbar\lambda[e^{i\omega t}(|1\rangle\langle 0|+|2\rangle\langle 0|)+e^{-i\omega t}(|0\rangle\langle 1|+|0\rangle\langle 2|)].
\end{align}
Interaction of the system with the bath is given by the same Liouvillian in Eq. (\ref{eq:Lc}). However the hot bath now is more tricky as it contains coherence between $1$ and $2$. Using Born-Markov and Wigner Weisskopf approximations, we obtain
\begin{widetext}
	\begin{align}\label{eq:Lh1}
		-\mathcal{L}_h[\rho_R]&=\Gamma_{h1}([|1\rangle\langle 1|[(n_{h1}+1)\rho_R-n_{h1}\rho_{gg}]+[n_{h1}\rho_R-(n_{h1}+1)\rho_{11}]|g\rangle\langle g|)\notag\\
		&+p\sqrt{\Gamma_{h1}\Gamma_{h2}}(|1\rangle\langle 2|[(n_{h2}+1)\rho_R-n_{h2}\rho_{gg}]-(n_{h2}+1)\rho_{21}|g\rangle\langle g|)\notag\\
		&+p\sqrt{\Gamma_{h1}\Gamma_{h2}}(|2\rangle\langle 1|[(n_{h1}+1)\rho_R-n_{h1}\rho_{gg}]-(n_{h1}+1)\rho_{12}|g\rangle\langle g|)\notag\\
		&+\Gamma_{h2}([|2\rangle\langle 2|[(n_{h2}+1)\rho_R-n_{h2}\rho_{gg}]+[n_{h2}\rho_R-(n_{h2}+1)\rho_{22}]|g\rangle\langle g|)\notag\\
		&+\Gamma_{h1}(|g\rangle\langle g|[n_{h1}\rho_R-(n_{h1}+1)\rho_{11}]+[(n_{h1}+1)\rho_R-n_{h1}\rho_{gg}]|1\rangle\langle 1|)\notag\\
		&+p\sqrt{\Gamma_{h1}\Gamma_{h2}}([(n_{h2}+1)\rho_R-n_h\rho_{gg}]|2\rangle\langle 1|-(n_{h2}+1)\rho_{12}|g\rangle\langle g|)\notag\\
		&+p\sqrt{\Gamma_{h1}\Gamma_{2h}}([(n_{h1}+1)\rho_R-n_{h1}\rho_{gg}]|1\rangle\langle 2|-(n_{h1}+1)\rho_{21}|g\rangle\langle g|)\notag\\
		&+\Gamma_{h2}(|g\rangle\langle g|[n_{h2}\rho_R-(n_{h2}+1)\rho_{22}]+[(n_{h2}+1)\rho_R-n_{h2}\rho_{gg}]|2\rangle\langle 2|),
	\end{align}
\end{widetext}
where $p=\frac{\mathbf{\mu_{1g}}\cdot\mathbf{\mu_{2g}}}{|\mathbf{\mu_{1g}}||\mathbf{\mu_{2g}}|}$ is the dipole alignment factor that characterizes the strength of noise induced coherence. Here we assume that states $1$ and $2$ are close so the coupling constant that enters $\Gamma_{1,2h}$ is flat.  However we keep explicitly the distinction between $n_{h1}$ and $n_{h2}$. We next assume that hot bath energy is tuned midway, which has been proven to have the strongest effect in lasers, as 
\begin{align}\label{eq:nh0}
	n_{h1,2}=\left[\exp\left(\frac{\hbar(\omega_{h}\pm \Delta)}{k_BT_h}\right)-1\right]^{-1}.
\end{align}

The lasing part of the Liouvillian is given by 
\begin{align}
	L_l[\rho_R]=-\frac{i}{\hbar}[H_0-\bar{\bar{H}}+V_R,\rho_R],
\end{align}
where 
\begin{align}
	\bar{\bar{H}}=\hbar\omega_g|g\rangle\langle g|+\frac{\hbar\omega}{2}(|1\rangle\langle 1|+|2\rangle\langle 2|-|0\rangle\langle 0|),
\end{align}
and $V_R$ is given by Eq. (\ref{eq:VR2}).
The time evolution of the density matrix equations is given by
\begin{align}
	\dot{\rho}_{11}&=i\lambda(\rho_{10}-\rho_{01})-2\Gamma_{h1}[(n_{h1}+1)\rho_{11}-n_{h1}\rho_{gg}]\notag\\
	&-p\sqrt{\Gamma_{h1}\Gamma_{h2}}(n_{h2}+1)[\rho_{12}+\rho_{21}],
\end{align}
\begin{align}
	\dot{\rho}_{22}&=i\lambda(\rho_{20}-\rho_{02})-2\Gamma_{h2}[(n_{h2}+1)\rho_{11}-n_{h2}\rho_{gg}]\notag\\
	&-p\sqrt{\Gamma_{h1}\Gamma_{h2}}(n_{h1}+1)[\rho_{12}+\rho_{21}],
\end{align}
\begin{align}
	\dot{\rho}_{00}&=i\lambda(\rho_{01}+\rho_{02}-\rho_{10}-\rho_{20})\notag\\
	&-2\Gamma_c[(n_c+1)\rho_{00}-n_c\rho_{gg}],
\end{align}
\begin{align}
	\rho_{gg}=1-\rho_{11}-\rho_{22}-\rho_{00},
\end{align}
\begin{align}
	\dot{\rho}_{12}&=-[i\omega_{12}+[\Gamma_{1h}(n_{h1}+1)+\Gamma_{h2}(n_{h2}+1)+\gamma_{12}]\rho_{12}\notag\\
	&+i\lambda(\rho_{10}-\rho_{02})-p\sqrt{\Gamma_{h1}\Gamma_{h2}}[(n_{h1}+1)\rho_{11}\notag\\
	&+(n_{h2}+1)\rho_{22}-(n_{h1}+n_{h2})\rho_{gg}],
\end{align}
\begin{align}\label{eq:r10}
	\dot{\rho}_{10}&=-[i(\omega_{10}-\omega)+\Gamma_c(n_c+1)+\Gamma_{h1}(n_{h1}+1)]\rho_{10}\notag\\
	&+i\lambda(\rho_{11}-\rho_{00}+\rho_{12})-p\sqrt{\Gamma_{h1}\Gamma_{h2}}(n_{h2}+1)\rho_{20},
\end{align}
\begin{align}
	\dot{\rho}_{20}&=-[i(\omega_{20}-\omega)+\Gamma_c(n_c+1)+\Gamma_{h2}(n_{h2}+1)]\rho_{20}\notag\\
	&+i\lambda(\rho_{22}-\rho_{00}+\rho_{21})-p\sqrt{\Gamma_{h1}\Gamma_{h2}}(n_{h1}+1)\rho_{10}.
\end{align}

The power, heat flux and efficiency of this QHE is given by
\begin{align}\label{eq:PQ}
	P_Q=-i\hbar\lambda[\omega_{10}(\rho_{01}-\rho_{10})+\omega_{20}(\rho_{02}-\rho_{20})],
\end{align}
\begin{align}
	\dot{Q}_{hQ}=i\hbar\lambda[\omega_{1g}(\rho_{01}-\rho_{10})+\omega_{2g}(\rho_{02}-\rho_{20})],
\end{align}
\begin{align}
	\eta_Q=-\frac{P_Q}{\dot{Q}_{hQ}}=\frac{\omega_{10}(\rho_{01}-\rho_{10})+\omega_{20}(\rho_{02}-\rho_{20})}{\omega_{1g}(\rho_{01}-\rho_{10})+\omega_{2g}(\rho_{02}-\rho_{20})},
\end{align}
where subscript $Q$ signifies quantum enhancement. Assuming that the laser frequency is tuned midway between states $1$ and $2$, $\omega=(\omega_{10}+\omega_{20})/2=\omega_{10}-\Delta=\omega_{20}+\Delta$, the efficiency can be recast as Eq. (\ref{eq:etQ}).

\subsection{Lower bound. Fixed $\omega_h$}

After solving the density matrix equations and substituting the solution into the expressions for power in (\ref{eq:PQ}) and efficiency in (\ref{eq:etQ}), neglecting dephasing $\gamma_{12}$, using strong coupling limit $\lambda\gg \Gamma_{h,c}$ in the high temperature limit $n_{h1,2}=\frac{T_h}{\omega_h\pm\Delta}\gg 1$, $n_c=\frac{T_c}{\omega_c}\gg 1$, assuming degenerate states ($\Delta=0$) and introducing $\gamma_p$ we obtain for the power
\begin{align}
	P_{Q}^{(\omega_h)}=\frac{(1-c)\gamma_p(1-c\tau)}{2c\gamma_p+2c^2\tau}.
\end{align}
Optimizing the efficiency with respect to $c$ we obtain Eq. (\ref{eq:etssdwh})  with $\gamma\to\gamma_p$.

\subsection{Upper bound. Fixed $\omega_c$}

For the power we obtain
\begin{align}
	P_{Q}^{(\omega_c)}=\frac{(1-c)\gamma_p(1-c\tau)}{2\gamma_p+2c\tau}.
\end{align}
Optimizing the power with respect to $c$, we obtain for the efficiency in Eq. (\ref{eq:etssdwc})  with $\gamma\to\gamma_p$.
If we include nondegenerate case such that $\Delta\neq 0$, the efficiency is reduced compared to its degenerate case.

\section{Low temperature operation}\label{sec:powerlow}

Although the high temperature limit yields the highest power an interesting physics especially in the context of coherence. We have previously demonstrated that the QHE power can be enhanced due to coherence. The maximum effect of doubling the power has been achieved at low temperature, $n_c=e^{-\hbar\omega_c/(k_BT_c)}$ and $n_h=e^{-\hbar\omega_h/(k_BT_h)}$. The power in this case is given by 
\begin{align}\label{eq:PlowT3lev}
	P_{SSD}^{(\omega_c)}=-\frac{2\gamma}{\gamma+1}(\omega_h-\omega_c)(n_h-n_c){\color{red}.}
\end{align}
Similarly for the four-level system we obtain
\begin{align}\label{eq:Plow4lev}
	P_{Q}=-\frac{\gamma_p}{1+\gamma_p}(\omega_h-\omega_c)(n_h-n_c).
\end{align}
Both expressions yield the same optimized efficiency, which is independent of $\gamma$ or $\gamma_p$, whereas power itself depends on $p$, which is discussed in the main text.


%

\end{document}